\documentclass[aps,pra,twocolumn,showpacs]{revtex4}

\usepackage{graphicx}

\begin{document}

\title{Improving entanglement concentration of Gaussian states by local displacements}

\author{Jarom\'{\i}r Fiur\'{a}\v{s}ek} 
\affiliation{Department of Optics, Palack\'{y} University, 17. listopadu 12,
77146 Olomouc, Czech Republic}

\begin{abstract}
We investigate entanglement concentration of  continuous-variable Gaussian states by local single-photon subtractions combined with local Gaussian operations.
We first analyze the local squeezing-enhanced entanglement concentration protocol proposed very recently by Zhang and van Loock [e-print: arXiv:1103.4500 (2011)] 
and discuss the mechanism by which local squeezing before photon subtraction helps to increase the entanglement of the output state of the protocol. 
We next show that a similar entanglement improvement 
can be achieved by using local coherent displacements instead of single-mode squeezing. 
\end{abstract}

\pacs{03.67.Mn,03.67.Hk, 42.50.Dv}

\maketitle 

\section{Introduction}

Distillation of continuous-variable entanglement is fundamentally limited by a no-go theorem which states
that experimentally accessible entangled Gaussian states cannot be distilled by local Gaussian operations and classical communication \cite{Eisert02,Giedke02,Fiurasek02}.
In order to increase entanglement of a Gaussian state such as two-mode squeezed vacuum by local operations, one thus has to resort to
non-Gaussian operations \cite{Duan00}. A prominent example of such an operation is the photon subtraction, that has been proposed for continuous-variable 
entanglement concentration in a seminal paper by Opatrn\'{y} \emph{et al.} \cite{Opatrny00}. Since then, the photon subtraction has been recognized as a crucial tool
for implementing many tasks in optical continuous-variable quantum information processing. Besides entanglement concentration \cite{Cochrane02,Olivares03,Ourjoumtsev07,Takahashi10,Zhang10}, 
this includes generation of Schr\"{o}dinger cat-like states by subtracting photons from single-mode squeezed vacuum state \cite{Ourjoumtsev06,Nielsen06,Wakui07,Gerrits10}, 
generation of arbitrary single-mode states of light beams \cite{Fiurasek05,Neergaard10}, implementation of quantum gates for qubits encoded into superpositions of two coherent states
\cite{Marek10}, and probabilistic noiseless amplification of coherent states of light \cite{Ralph08,Marek10b,Fiurasek09,Xiang10,Ferreyrol10,Usuga10,Zavatta11}.

Although Gaussian operations alone are useless for distillation of Gaussian entanglement, it has been recently shown by Zhang and van Loock \cite{Zhang11}
that the performance of entanglement concentration via local single-photon subtraction can be enhanced  by suitable local
squeezing operations. In this paper we further investigate this intriguing concept and clarify the origin of enhancement of entanglement concentration
due to the local pre-squeezing. We next show that the same effect can be achieved also by local displacements, which are much easier and experimentally more feasible operations than
squeezing. Indeed, the combination of photon subtraction and coherent displacement has been recently successfully experimentally utilized to generate 
arbitrary superpositions of squeezed vacuum and squeezed single-photon states from squeezed vacuum \cite{Neergaard10}. 
Coherent displacement can be implemented by mixing the signal beam with a weak coherent laser beam on a highly unbalanced beam splitter.
In contrast, squeezing is much more experimentally challenging and resource demanding operation. Squeezing in cavity-based optical parametric amplifier 
would require efficient injection of the pulsed signal beam in the cavity as well as efficient retrieval of the squeezed beam which is a nontrivial task.
Squeezing based on a single passage through a nonlinear crystal pumped by a strong pulsed laser still requires careful modematching and may suffer from noise due to
 spontaneous parametric down-conversion producing pairs of photons in many modes. Alternatively, one can employ an off-line generated
 auxiliary squeezed vacuum state combined with interference on an unbalanced beam splitter, homodyne detection and feedforward \cite{Filip05,Yoshikawa07}. 
This squeezing technique was successfully experimentally tested,  however, the operation is imperfect and inevitably adds some extra noise
due to finite squeezing of the auxiliary state.

The rest of the paper is structured as follows. In Section II we present a simple analysis of the local squeezing-enhanced entanglement concentration 
which reveals the origin of the entanglement enhancement in this scheme. In Section III we describe the displacement-enhnaced entanglement concentration procedure
and we analytically determine the optimal displacement and success probability scaling in the limit of pure weakly squeezed input states,
where the scheme achieves the maximum entanglement enhancement. We also numerically calculate the optimal local displacements maximizing entanglement 
of the distilled state for both pure and mixed Gaussian two-mode squeezed states. Finally, Section IV contains brief summary and  conclusions.

\begin{figure}[t]
\includegraphics[width=0.8\linewidth]{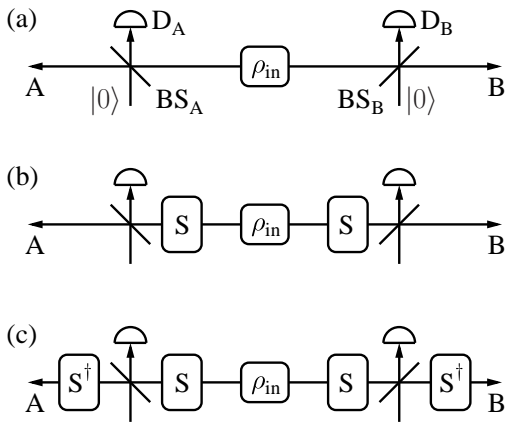}
\caption{(a) Entanglement concentration by single-photon subtraction. BS$_A$ and BS$_B$ are highly unbalanced beam splitters, $\hat{\rho}_{\mathrm{in}}$ is the shared (generally mixed) entangled state,
and D$_A$ and D$_B$ denote single-photon detectors. (b) Entanglement concentration improved by local single-mode suqeezing $\hat{S}$. The state can be un-squeezed after concentration 
by inverse local operations $\hat{S} ^\dagger$ without changing its entanglement (c).}
\end{figure}

\section{Squeezing-enhanced entanglement concentration}

The considered entanglement concentration scheme is depicted in Fig. 1(a). The entangled two-mode squeezed state is distributed from a source
via some quantum channel to the two recipients traditionally called Alice and Bob who attempt to conditionally increase  the entanglement of the shared state
by local photon subtraction, which is a specific example of a quantum filter. The whole scheme is thus a typical example of a Procrustean entanglement concentration
protocol \cite{Bennett96}. In practice, the photon subtraction is accomplished by means of a highly  unbalanced beam splitter that  reflects a tiny part of the incoming 
beam on a single-photon detector APD. A successful photon subtraction is heralded by a click of the detector. Alice and Bob exchange their measurement results via classical communication line and the 
entanglement concentration succeeds if they both register detector clicks. In the limit of very low beam splitter reflectance, the resulting operation can be described
by action of annihilation operators $\hat{a}$ and $\hat{b}$ on modes A and B, respectively, even if the detectors have non-unit detection efficiency and cannot 
distinguish number of photons.

It is instructive to first investigate concentration of a pure two-mode squeezed vacuum state
\begin{equation}
 |\psi\rangle=\sqrt{1-\lambda^2}\sum_{n=0}^\infty \lambda^n |nn\rangle,
\end{equation}
where $|n\rangle$ denotes Fock state, $|nn\rangle$ is a short-hand notation for $|n\rangle|n\rangle$, 
and $\lambda=\tanh r$, where $r$ stands for the two-mode squeezing constant. The non-normalized state after local photon subtractions
can be expressed as 
\begin{equation}
\hat{a}\otimes\hat{b}\,|\psi\rangle=\sqrt{1-\lambda^2}\sum_{n=0}^\infty (n+1)\lambda^{n+1} |nn\rangle.
\end{equation}
Alice and Bob may amend the entanglement concentration protocol by local squeezing of modes A nad B prior to the photon subtraction \cite{Zhang11}, c.f. Fig. 1(b).
The filtering operation then reads $\hat{a}\otimes\hat{b} \, \hat{S}_A(s)\otimes\hat{S}_B(s)$, where $\hat{S}_M$ is a single-mode squeezing operation on mode M with
squeezing constant $s$. In order to clarify why the pre-squeezing can be useful, we apply  an inverse unitary squeezing operation
$\hat{S}_A^\dagger(s)\otimes\hat{S}_B^\dagger(s)$ to the concentrated state which does not change its entanglement, see Fig. 1(c). 
The overall filtering operation $\hat{F}_{AB}$ then reads
\begin{equation}
\hat{F}_{AB}=\hat{S}_A^\dagger(s)\hat{a}\hat{S}_A(s)\otimes\hat{S}_B^\dagger(s)\hat{b} \hat{S}_B(s).
\end{equation}
The squeezing transforms the annihilation operators into linear combination of annihilation and creation operators, and we have
\begin{equation}
\hat{F}_{AB}= [\hat{a}\cosh(s)+\hat{a}^\dagger \sinh(s)]\otimes [\hat{b}\cosh(s)+\hat{b}^\dagger \sinh(s)].
\end{equation}
We can thus see that the local pre-squeezing effectively converts the photon subtraction into a coherent combination of photon subtraction and photon addition.

Let us now investigate what happens in the limit of weak two-mode squeezing, $\lambda\ll 1$. We can write
\begin{equation}
\hat{F}_{AB}|\psi\rangle =\hat{F}_{AB}\left(|00\rangle+\lambda|11\rangle\right)+O(\lambda^2),
\end{equation}
where $O(\lambda^2)$ indicates small terms of the order of $\lambda^2$.
After some algebra we get
\begin{eqnarray*}
\hat{F}_{AB}|\psi\rangle& =&\lambda\cosh^2(s) |00\rangle+\sinh^2(s) |11\rangle+2\lambda\sinh^2(s)|22\rangle \\
& &+\frac{\lambda}{\sqrt{2}}\sinh(2s)\left(|02\rangle+|20\rangle\right)+O(\lambda^2),
\end{eqnarray*}
If we set $\tanh(s)=\sqrt{\lambda}$, then we obtain
\begin{equation}
\hat{F}|\psi\rangle \approx \lambda (|00\rangle+|11\rangle) + O(\lambda^{3/2}).
\end{equation}
Note that the amplitude of the states $|02\rangle$ and $|20\rangle$ is suppressed by a factor of $\sqrt{\lambda}$ with respect
to the amplitude of the dominant term $|00\rangle+|11\rangle$. In the limit of very small $\lambda$ the result of entanglement concentration
is a maximally entangled state of two qubits hence the protocol can extract one e-bit of entanglement from arbitrarily weakly squeezed initial state.
The success probability of the protocol scales as $P_{\mathrm{succ}}\propto \lambda^2$ for $\lambda \ll 1$. The squeezing enhanced protocol should be contrasted 
with the ordinary concentration by local photon subtraction that in the limit $\lambda\ll 1$ provides only very weakly entangled state $|00\rangle+2\lambda|11\rangle$.

\begin{figure}[t]
\includegraphics[width=0.8\linewidth]{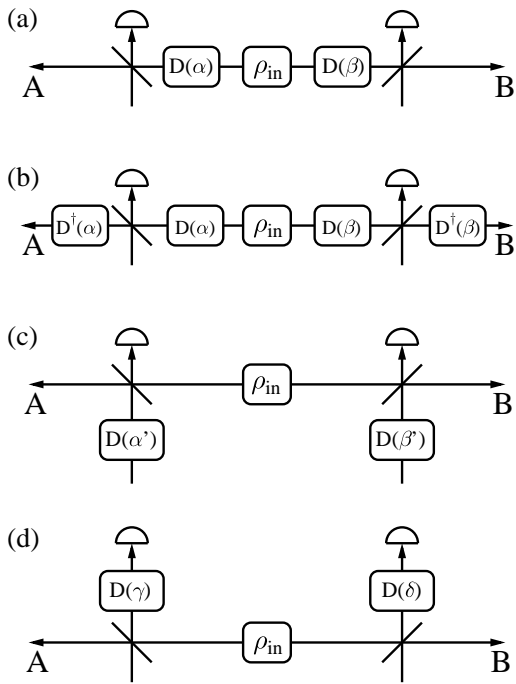}
\caption{Entanglement concentration scheme combining local displacements and photon subtraction (a). Applying local unitary 
inverse displacement operations on the state after concentration does not change its entanglement (b). Instead of displacing the signal modes one can equivalently inject weak coherent states
into the auxiliary ports of the unbalanced beam splitters (c) or displace the beams impinging on single-photon detectors (d).}
\end{figure}

\section{Displacement-enhanced entanglement concentration}

We next show that similar improvement in the amount of extracted entanglement can be obtained if we replace the local squeezing operations with local 
displacements $\hat{D}_A(\alpha)$ and $\hat{D}_B(\beta)$, see Fig. 2(a). Similarly as above, we assume that the displacement is undone after photon subtraction by inverse displacements
$\hat{D}_A^\dagger(\alpha)$ and $\hat{D}_B^\dagger(\beta)$, c.f. Fig. 2(b). The resulting local quantum filter reads
\begin{equation}
\hat{G}_{AB}=\hat{D}_A^\dagger(\alpha)\hat{a}\hat{D}_A(\alpha)\otimes\hat{D}_B^\dagger(\beta)\hat{b} \hat{D}_B(\beta).
\end{equation}
The local displacements transform the photon subtraction into a coherent superposition of photon subtraction and identity operation and 
we have
\begin{equation}
\hat{G}_{AB}=(\hat{a}+\alpha)\otimes (\hat{b}+\beta).
\end{equation}
Let us again consider the limit $\lambda \ll 1$. We obtain
\begin{equation}
\hat{G}_{AB}|\psi\rangle= (\alpha\beta+\lambda)|00\rangle+\lambda\left(\alpha|10\rangle+\beta|01\rangle\right)+\lambda\alpha\beta |11\rangle+O(\lambda^2).
\label{Gpsi}
\end{equation}
The apparently dominant vacuum term corresponds to a separable state, but it can be completely eliminated by a destructive interference provided
that $\alpha\beta=-\lambda$. The maximal entanglement is then obtained if the displacements are real and symmetric,
\begin{equation}
\alpha=-\beta =\sqrt{\lambda}.
\label{ablambda}
\end{equation}
If we insert $\alpha$ and $\beta$ given by Eq. (\ref{ablambda}) into (\ref{Gpsi}) we get
\begin{equation}
\hat{G}_{AB}|\psi\rangle=\lambda^{3/2}(|10\rangle-|01\rangle)+O(\lambda^2).
\end{equation}
We can see that in the limit $\lambda \ll 1$ we again extract one e-bit of entanglement, although now in a slightly different form of a single
photon split among the two modes A and B. The success probability of the scheme scales as $P_{\mathrm{succ}}\propto \lambda^3$ for small $\lambda$.
In comparison with the pre-squeezing strategy the success rate is thus smaller by a factor of $\lambda$. Nonetheless, the simplicity of
displacement operation compensates for this reduced efficiency. 

Note that the local photon subtractions combined with displacements can be seen as a local version of the non-local coherent photon 
subtraction demonstrated by Ourjoumtsev \emph{et al.} \cite{Ourjoumtsev07}. There the two beams reflected off the beam splitters BS$_A$ and BS$_B$ were combined on another 
balanced beam splitter and a click of an APD placed on one of its output ports heralded coherent photon subtraction associated 
with the filter $\hat{G}_{AB,\mathrm{nl}}=\hat{a}+\hat{b}$.
Surprisingly, the present local scheme becomes equivalent to that non-local strategy in the limit of small $\lambda$ and for
carefully tuned local displacements (\ref{ablambda}). Note also that the displacements can be equivalently performed on the ancilla input ports,
e.g. by injecting weak coherent states there, see Fig. 2(c). Yet another possibility is to displace the reflected beam right before the single photon detection, c.f. Fig. 2(d).
This latter scheme has been recently successfully experimentally utilized to generate qubit states formed by squeezed
superpositions of vacuum and single-photon states \cite{Neergaard10}.

Let us now analyze the performance of the displacement-enhanced entanglement concentration scheme beyond the limit of $\lambda \ll 1$.
In our model we take into account a non-zero reflectance $R$ of the beam-splitters BS$_A$ and BS$_B$ and we assume on-off single-photon detectors
described by a two-component POVM $\hat{\Pi}_0=|0\rangle\langle 0|$ and $\hat{\Pi}_1= \hat{I}-\hat{\Pi}_0$, where $\hat{I}$ denotes
the identity operator and $\hat{\Pi}_1$ corresponds 
to a click of the the detector. For the sake of simplicity we assume that the detectors have unit efficiency. Experiments usually involve
highly unbalanced beam splitters with $R\ll 1$. In this limit, the main effect of non-unit detection efficiency $\eta$ is
 the reduction of success rate by a factor of $\eta^2$. More specifically,
the entanglement concentration scheme with inefficient detectors is equivalent to a scheme with perfect detectors and reduced 
reflectance of the beam splitters, $R'=\eta R/[1-(1-\eta) R]$, preceded by the transmission of the  state through a lossy channel with transmittance 
$\tilde{T}=1-(1-\eta)R$. In the limit of $R \ll 1$ the latter can be neglected since $\tilde{T} \approx 1$. We allow for an arbitrary initial mixed two-mode
Gaussian state $\hat{\rho}_{\mathrm{in}}$ as we shall investigate entanglement concentration of two-mode squeezed vacuum states transmitted over lossy channels.
We denote by $\hat{U}_{\mathrm{BS},AC}$ and $\hat{U}_{\mathrm{BS},BD}$ the unitary operation describing mixing of two modes on the unbalanced beam splitters 
with reflectance $R$. The non-normalized density matrix of the concentrated state can be expressed as follows \cite{Zhang11}
\begin{equation}
\hat{\rho}_{\mathrm{out}}= \mathrm{Tr}_{CD} \left[
(\hat{\Pi}_{CD}\otimes\hat{I}_{AB})\, \hat{U}\,(\hat{\rho}_{\mathrm{in},AB}\otimes |00\rangle\langle 00|_{CD})\,\hat{U}^\dagger\right],
\label{rhoout}
\end{equation}
where 
\begin{equation}
\hat{U}=\hat{U}_{\mathrm{BS},AC}\otimes \hat{U}_{\mathrm{BS},BD} \, \hat{D}_A(\alpha)\otimes \hat{D}_B(\beta)\otimes \hat{I}_{CD},
\end{equation}
$\hat{\Pi}_{CD}=\hat{\Pi}_{1,C}\otimes\hat{\Pi}_{1,D}$, and $\mathrm{Tr}_{CD}$ denotes a 
partial trace over modes C and D that are initially prepared in the vacuum state $|00\rangle_{CD}$.

We quantify the entanglement of the concentrated state $\hat{\rho}_{\mathrm{out}}$ by logarithmic negativity $E_{N}$, which is a computable measure of 
entanglement for non-Gaussian mixed states \cite{Vidal02}. Note that $\hat{\rho}_{\mathrm{out}}$ is mixed even for pure input $\hat{\rho}_{\mathrm{in}}$ due to the inability
of the APDs to resolve number of photons. Recall that 
\begin{equation}
E_N(\hat{\rho}_{AB})=\log_2 || \hat{\rho}_{AB}^{T_A} ||_1,
\end{equation}
where $T_A$ denotes partial transposition with respect to mode A  and $||\hat{X} ||_1=\mathrm{Tr}\sqrt{{\hat{X}}^\dagger \hat{X}}$.
In order to calculate $E_N$ we need to determine matrix elements of $\hat{\rho}_{\mathrm{out}}$ in Fock state basis.  One option
is to explore the fact that for a Gaussian input state $\hat{\rho}_{\mathrm{in}}$ the output state $\hat{\rho}_{\mathrm{out}}$ can be expressed as 
a linear combination of four Gaussian states \cite{Fiurasek05,Patron05,Zhang11}. 
For a Gaussian state, the density matrix elements in Fock basis can be expressed in terms of multivariate
Hermite polynomials because the Husimi Q-function of the state, which is a generating function of these matrix elements, has a Gaussian form.
 Here we follow a more straightforward route and directly
evaluate Eq. (\ref{rhoout}) in Fock basis truncated at Fock state $n_{\mathrm{max}}=10$. For the considered values of $\lambda$ and other parameters,
the errors due to the truncation are negligible.

\begin{figure}[t]
\includegraphics[width=0.99\linewidth]{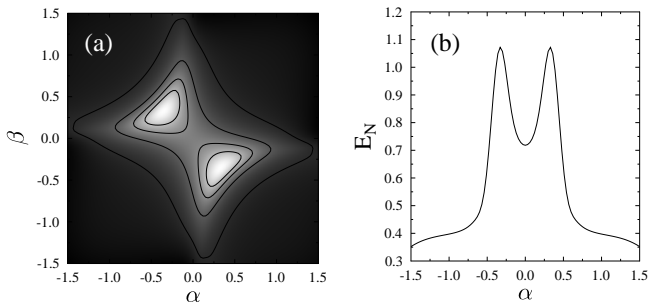}
\caption{Logarithmic negativity $E_N$ of two-mode squeezed state after entanglement concentration. Contour plot in panel (a) shows 
dependence of $E_N$ on the local displacements  $\alpha$ and $\beta$ and panel 
(b) represents the cut through panel (a) along the diagonal line $\beta=-\alpha$.  The parameters read $\lambda=0.15$ and $R=0.1$.}
\end{figure}

In Fig. 3 we plot the dependence of logarithmic negativity of concentrated state on the local displacements $\alpha$ and $\beta$
for initial pure two-mode squeezed vacuum with $\lambda=0.15$ and beam splitters with $10\%$ reflectance, $R=0.1$.  We can see that
the two peaks of maximum entanglement lie on the line $\alpha=-\beta$, as expected from the symmetry. 
Numerical calculations indicate that it suffices to consider real displacements, adding complex phase factors does not further increase the entanglement. 
In the rest of the paper we thus restrict ourselves to real $\alpha$ and always set $\beta=-\alpha$.

 The dependence of the performance of entanglement concentration protocol on the two-mode squeezing constant $\lambda$ is illustrated in Fig. 4.
The logarithmic negativities of an initial pure two-mode squeezed vacuum state, a state after local single-photon subtractions, and a state
after local single-photon subtractions combined with optimal local displacements are shown in Fig. 4(a). We can see that the scheme
with local displacements outperforms the scheme involving only photon subtractions and it yields almost $1$ e-bit of entanglement even
in the limit $\lambda \ll 1$. The slight reduction of $E_N$ below $1$ e-bit is caused by on-off single-photon detectors
leading to mixed output state after entanglement concentration. The data plotted in Fig. 4(b) confirm that 
the success probability of the protocol scales as $R^2\lambda^3$ (with displacements) and $R^2 \lambda^2$ (without displacements).
The optimal displacement $\alpha_{\mathrm{opt}}$ 
follows the scaling $\alpha_{\mathrm{opt}}=\sqrt{\lambda}$ for small $\lambda$, c.f. Fig. 4(c), but for larger $\lambda$ it holds that 
$\alpha_{\mathrm{opt}} < \sqrt{\lambda}$.

\begin{figure}[b]
\includegraphics[width=0.99\linewidth]{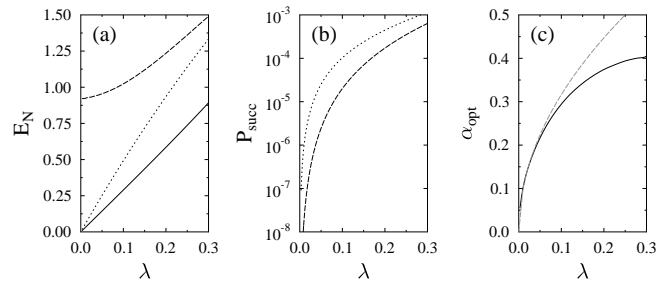}
\caption{Performance of entanglement concentration in dependence on the two-mode squeezing constant $\lambda$. In panel (a) we plot the
logarithmic negativities $E_N$ of three states: initial pure two-mode squeezed vacuum (solid line), a two-mode state after local photon subtractions (dotted line) 
and a two-mode state after local photon subtractions combined with optimal local displacements (dashed line). Panel (b) shows the success rates $P_{\mathrm{succ}}$ of the concentration 
protocol with (dashed line) and without (dotted line) displacements. Panel (c) shows the optimal displacement $\alpha_{\mathrm{opt}}$ (solid black line)
and the $\sqrt{\lambda}$ reference curve  for comparison  (gray dashed line). The results were obtained for $R=0.1$.}
\end{figure}

Finally we investigate entanglement concentration of mixed states obtained by transmitting each mode of a pure two-mode squeezed vacuum state
through a lossy channel with transmittance $1-\nu$. Results of numerical simulations plotted in Fig. 5 reveal that the local displacements
help to increase the attainable entanglement even in the presence of losses although the achievable entanglement decreases with increasing channel
losses $\nu$. Also the success probability of concentration and the optimal displacement decrease with increasing losses, see Fig. 5(b) and 5(c).
This effect can be attributed to signal attenuation and reduction of mean number of photons due to transmission through a lossy channel.

\begin{figure}[t]
\includegraphics[width=0.99\linewidth]{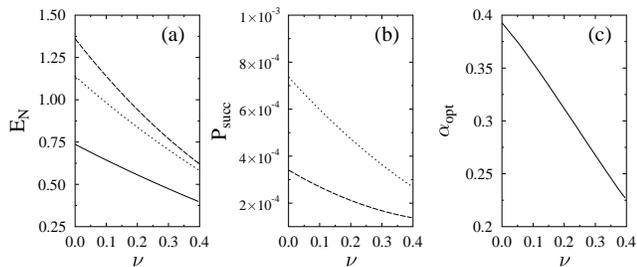}
\caption{Entanglement concentration of a two-mode squeezed state transmitted over a lossy channel with loss factor $\nu$.
Panel (a) shows the logarithmic negativities $E_N$ of an initial mixed two-mode squeezed state (solid line), a two-mode state after local photon subtractions (dotted line) 
and a two-mode state after local photon subtractions combined with optimal local displacements (dashed line). Panel (b) shows the success rates $P_{\mathrm{succ}}$ of the concentration 
protocol with (dashed line) and without (dotted line) displacements. Panel (c) shows the optimal displacement $\alpha_{\mathrm{opt}}$.
The other parameters read $\lambda=0.25$ and $R=0.1$.}
\end{figure}

\section{Conclusions}
In summary we have investigated continuous-variable entanglement concentration schemes based on local photon subtractions combined with
local Gaussian operations. A simple analysis in Fock basis provided an insight into the working principle of the squeezing-enhanced entanglement 
concentration scheme proposed by Zhang and van Loock \cite{Zhang11}. We have next shown that a similar entanglement enhancement can be achieved 
by employing local displacements
 instead of single-mode squeezing. Remarkably, the entanglement is enhanced by a destructive quantum interference that suppresses or even
 completely eliminates the otherwise dominant vacuum term. The displacement-based scheme is much easier to implement experimentally than 
 the scheme with local squeezers albeit at the cost of a somewhat reduced success rate of the protocol. Note also that while the entanglement concentration 
 schemes involving local displacement or local squeezing can produce similar amount of entanglement, the internal structure of the entangled states
 is rather different. In the weak-squeezing limit and up to local Gaussian operations the displacement enhanced scheme essentially yields an entangled single-photon state $\frac{1}{\sqrt{2}}(|10\rangle-|01\rangle)$
  while the scheme with local squeezers produces a state exhibiting almost perfect photon 
 number correlations $\frac{1}{\sqrt{2}}(|00\rangle+|11\rangle)$.

 The analytical and numerical results indicate that the entanglement enhancement 
achieved by single photon subtractions combined with optimal local 
displacements is of the order of 1 e-bit. Here the entanglement enhancement is defined as a difference between entanglement of concentrated and input states. 
Even higher entanglement enhancement could be achieved by more sophisticated local filtering operations
such as the recently proposed probabilistic noiseless amplifier that approximates the unphysical operation $g^{\hat{n}}$, where $g>1$
is the amplification gain \cite{Ralph08,Marek10b,Fiurasek09,Xiang10,Ferreyrol10,Usuga10,Zavatta11}. Such amplifier may in principle generate arbitrarily large amount of entanglement from any initial pure two-mode squeezed vacuum 
state. However, the resulting scheme may become rather complex and its success rate very small.

 From a more general perspective, the entanglement concentration protocol analyzed in this paper is an example of a CV quantum information
 processing scheme whose performance can be significantly enhanced by a suitable preprocessing with Gaussian operations. Another example of such 
 improvement is the recently proposed interface between two weakly coupled quantum systems \cite{Filip09,Marek10c} where a perfect transmission of quantum 
 state from one system to the other can be achieved with the help of ancilla modes and auxiliary Gaussian operations.
 It will be interesting to investigate whether similar Gaussian pre- or post-processing may be beneficial also in other CV QIP protocols.

\emph{Note added:} After this work was completed we became aware of a related paper
by Lee \emph{et al.} \cite{Lee11} where entanglement concentration of two-mode squeezed vacuum 
by linear combination of photon addition and subtraction is investigated. This protocol is equivalent to the protocol 
discussed in Sec. II of the present paper, and it is shown in Ref. \cite{Lee11} that one ebit 
of entanglement can be obtained by this method in the limit of weak squeezing.

\acknowledgments
We acknowledge the financial support of the Future and Emerging Technologies (FET) programme within the Seventh Framework Programme
for Research of the European Commission, under the FET-Open grant agreement COMPAS, number 212008, co-financed by the Czech Ministry of Education (7E08028).
This work was also supported by the Czech Ministry of Education under the projects Center of Modern Optics (LC06007) and Measurement and Information in Optics (MSM6198959213).

\end{document}